\documentclass[%
reprint,
superscriptaddress,
showpacs,preprintnumbers,
 amsmath,amssymb,
aps,
prr,
floatfix
]{revtex4-1}
\usepackage{amsmath}
\usepackage{graphicx,times}
\DeclareMathSizes{12}{17.28}{9}{7}

\usepackage{float}
\usepackage{dcolumn}
\usepackage{bm}
\usepackage{bbold}
\usepackage[multidot]{grffile}
\usepackage[colorlinks=true, breaklinks=true]{hyperref}
\usepackage[mathlines]{lineno}

\usepackage{comment}
\usepackage{xcolor}

\usepackage{xspace}

\usepackage[utf8]{inputenc}
\usepackage{scalerel,tikz}
\usetikzlibrary{shapes.geometric,fit,arrows,babel}
\usepackage[utf8]{inputenc} 
\usepackage[T1]{fontenc}    
\hypersetup{colorlinks=true,linkcolor=blue,citecolor=blue,urlcolor=blue}
\usepackage{url}            
\usepackage{booktabs}       
\usepackage{amsfonts}       
\usepackage{nicefrac}       
\usepackage{microtype}      
\usepackage{lipsum}
\usepackage{amsmath,amssymb,amsthm}
\usepackage[normalem]{ulem}
\usepackage{times}
\usepackage[colorinlistoftodos,prependcaption,textsize=tiny]{todonotes}
\usepackage{graphicx}
\usepackage{wrapfig}

\definecolor{Agreen}{rgb}{0.1, 0.6, 0.1} 
\usepackage{ulem}
\usetikzlibrary{svg.path}
\definecolor{orcidlogocol}{HTML}{A6CE39}
\tikzset{
	orcidlogo/.pic={
		\fill[orcidlogocol] svg{M256,128c0,70.7-57.3,128-128,128C57.3,256,0,198.7,0,128C0,57.3,57.3,0,128,0C198.7,0,256,57.3,256,128z};
		\fill[white] svg{M86.3,186.2H70.9V79.1h15.4v48.4V186.2z}
		svg{M108.9,79.1h41.6c39.6,0,57,28.3,57,53.6c0,27.5-21.5,53.6-56.8,53.6h-41.8V79.1z M124.3,172.4h24.5c34.9,0,42.9-26.5,42.9-39.7c0-21.5-13.7-39.7-43.7-39.7h-23.7V172.4z}
		svg{M88.7,56.8c0,5.5-4.5,10.1-10.1,10.1c-5.6,0-10.1-4.6-10.1-10.1c0-5.6,4.5-10.1,10.1-10.1C84.2,46.7,88.7,51.3,88.7,56.8z};}}
\newcommand\orcid[1]{\href{https://orcid.org/#1}{\mbox{\scalerel*{\begin{tikzpicture}[yscale=-1,transform shape]\pic{orcidlogo};\end{tikzpicture}}{|}}}}
\begin{document}

\title{Optimizing and reducing stochastic resonance by noise color in globally coupled bistable systems }
\author{Cong Liu\orcid{0000-0003-1705-4438}}
\email{Contact author: cliu@hebtu.edu.cn}
\affiliation{ College of Physics and Hebei Key Laboratory of Photophysics Research and Application, Hebei Normal University, Shijiazhuang, Hebei 050024, China }%
\affiliation{%
Lanzhou Center for Theoretical Physics, Key Laboratory of Theoretical Physics of Gansu Province, and Key Laboratory of Quantum Theory and Applications of MoE, Lanzhou University, Lanzhou, Gansu 730000, China
}%
\affiliation{Institute of Computational Physics and Complex Systems,
Lanzhou University, Lanzhou, Gansu 730000, China}
\author{Xin-Ze Song\orcid{0009-0005-9343-2391}}
\email{aaasongxinze@163.com}
\affiliation{%
Lanzhou Center for Theoretical Physics, Key Laboratory of Theoretical Physics of Gansu Province, and Key Laboratory of Quantum Theory and Applications of MoE, Lanzhou University, Lanzhou, Gansu 730000, China
}%
\affiliation{Institute of Computational Physics and Complex Systems,
Lanzhou University, Lanzhou, Gansu 730000, China}
\author{Zhi-Xi Wu\orcid{0000-0002-2982-2177}}%
\email{wuzhx@lzu.edu.cn}
\affiliation{%
Lanzhou Center for Theoretical Physics, Key Laboratory of Theoretical Physics of Gansu Province, and Key Laboratory of Quantum Theory and Applications of MoE, Lanzhou University, Lanzhou, Gansu 730000, China
}%
\affiliation{Institute of Computational Physics and Complex Systems, Lanzhou University, Lanzhou, Gansu 730000, China}

\author{Guo-Yong Yuan\orcid{0000-0002-8556-1303}}%
\affiliation{ College of Physics and Hebei Key Laboratory of Photophysics Research and Application, Hebei Normal University, Shijiazhuang, Hebei 050024, China }%

\date{\today}

\begin{abstract}
We investigate the collective signal response of two typical nonlinear dynamical models, the mean-field coupled overdamped bistable oscillators and the underdamped Duffing oscillators, with respect to both the additive Ornstein-Uhlenbeck noise and the weak periodical stimulus. 
Based on the linear response theory, we theoretically derive the dependences of the ensemble signal response on the noise intensity and driving frequency of both systems. 
Furthermore, we theoretically demonstrate that the noise color monotonically weakens the strength of stochastic resonance in the overdamped situation, but nonmonotonically strengthens it in the underdamped counterpart. Such a result goes against the conventional wisdom that the color of the additive noise impairs the magnitude of stochastic resonance. 
Finally, we perform the numerical integration to verify our theoretical results and discuss potential connections with the functional significance of $1/f$ noise. 
\end{abstract}

\maketitle


\emph{Introduction.}--Noise-driven nonlinear dynamical systems out of equilibrium are ubiquitous in nature. As striking examples, the annual fluctuations in solar radiation are crucial to the periodic changes in Earth's climate~\cite{Lucarini2019,Ghil2020}. A noisy environment promotes species coexistence in nature~\cite{Lai2005pre,Lai2005prl}. Pollen particles suspended in the fluid wander randomly due to the fluctuating force~\cite{Hanggi1990,Lucarini2019pre}, etc. Among these distinct scenes, a somewhat counterintuitive phenomenon stochastic resonance (SR), in which an intermediate level of noise can maximize the response of a system to a weak periodic stimuli, has been successively substantiated~\cite{Lai2005pre,Lai2005prl,Hanggi1990,Lucarini2019pre,Benzi1982,Nicolis1982,Gammaitoni1998}. Despite the notion of SR was originally designed to address the issue of the periodically recurrent ice ages~\cite{Benzi1982,Nicolis1982}, nowadays, it has been extended to almost all kinds of disciplines~\cite{Peters2021}, see Refs ~\cite{Gammaitoni1998} for a comprehensive review.

The dynamical archetype for SR is the overdamped small particle moving in a symmetrical double-well potential~\cite{Gammaitoni1998,Lucarini2019pre,Jung1991}. Since the fluctuation occurs on the time scale which is much smaller than that of the particle, the correlation function of the fluctuating force can be regarded ideally as a $\delta$-function, namely the fluctuating force can be represented by the Gaussian white noise~\cite{Hanggi1994}. Such a random force evokes a transition between the neighboring fixed points. Additionally, based on the Kramers escape theory~\cite{Hanggi1990,Lucarini2019pre}, the noise-induced average transition rate increases exponentially with the rising of the noise intensity, and when such a transition rate matches the external driving force, i.e., transition rate is twice the external driving frequency, the maximum amplitude of average oscillation can be observed~\cite{Gammaitoni1998,Lucarini2019pre}. 

Nevertheless, the ideal treatment of Gaussian white noise is inaccessible to an experimentalist or the physical world since the generation of white noise requires an infinite amount of power~\cite{Hanggi1994,Hanggi1993}. On the other hand, the colored noise concerning a finite correlation time plays a cornerstone role in physics~\cite{Hanggi1994,Hanggi1993,Baldassarri2002,Jung1988}, biology~\cite{Struzik2004,Kamenev2008,Peng1993}, engineering~\cite{Zhang2010,Mondal2018}, and neuroscience~\cite{Nozaki1999prl,Nozaki1999pre,Soma2003}, and is theorized to be a characteristic signature of complexity~\cite{Gilden1995}. For example, colored noises are associated with certain basic aspects of human cognition~\cite{Gilden1995}. In consequence, a great deal of interest has been devoted to the study of SR induced by colored noise. For instance, H\"{a}nggi and coworkers theoretically demonstrated that the noise color (correlation time) weakens the magnitude of SR for a single overdamped bistable oscillator driven by the additive colored noise~\cite{Hanggi1993}. Later on, Nozaki and collaborators verified such a conclusion through both the \textit{in vitro} neuronal sensory experiment of the rat skin~\cite{Nozaki1999prl,Nozaki1999pre} and the theoretical reduced model. These results established a conventional wisdom that the color of the additive
noise impairs the magnitude of stochastic resonance~\cite{Mondal2018,Soma2003,Gammaitoni1989,Cabrera1999}. Besides, the interest of colored noise-modulated SR has been also extended to the coupled or extended multiparticle systems. For example, Lorenzo and coworkers numerically revealed that a resonance effect between the chaotic attractor time scale and the noise correlation time can be observed in coupled Lorenz cells~\cite{Lorenzo1999}. Yang \emph{et al} investigated SR and synchronization induced by colored noise in globally coupled linear oscillators~\cite{Yang2016}. Particularly, based on the approximation of moments, Kang and collaborators proposed a semi-analytic method for quantitatively investigating the long-time ensemble dynamics of the mean-field coupled bistable oscillators~\cite{Kang2008}, etc. Despite these findings, whether the colored noise outperforms the white noise in evoking the collective signal response in coupled systems remains unclear.

In this letter, based on the linear response theory, we theoretically investigate the phenomenon of SR driven by the Ornstein-Uhlenbeck (OU) noise, a prototype of colored noise, in globally coupled overdamped bistable and underdamped Duffing oscillators. We give the dependences of the ensemble signal response on noise intensity and the driving frequency, respectively. Additionally, we theoretically demonstrate that the noise color monotonically impairs the strength of SR for overdamped bistable oscillators but nonmonotonically strengthens it in the underdamped counterpart. We perform numerical experiments to verify our theoretical results and discuss their potential applications with the functional significance of colored noise.


\begin{figure}
\includegraphics[width=1.0\linewidth]{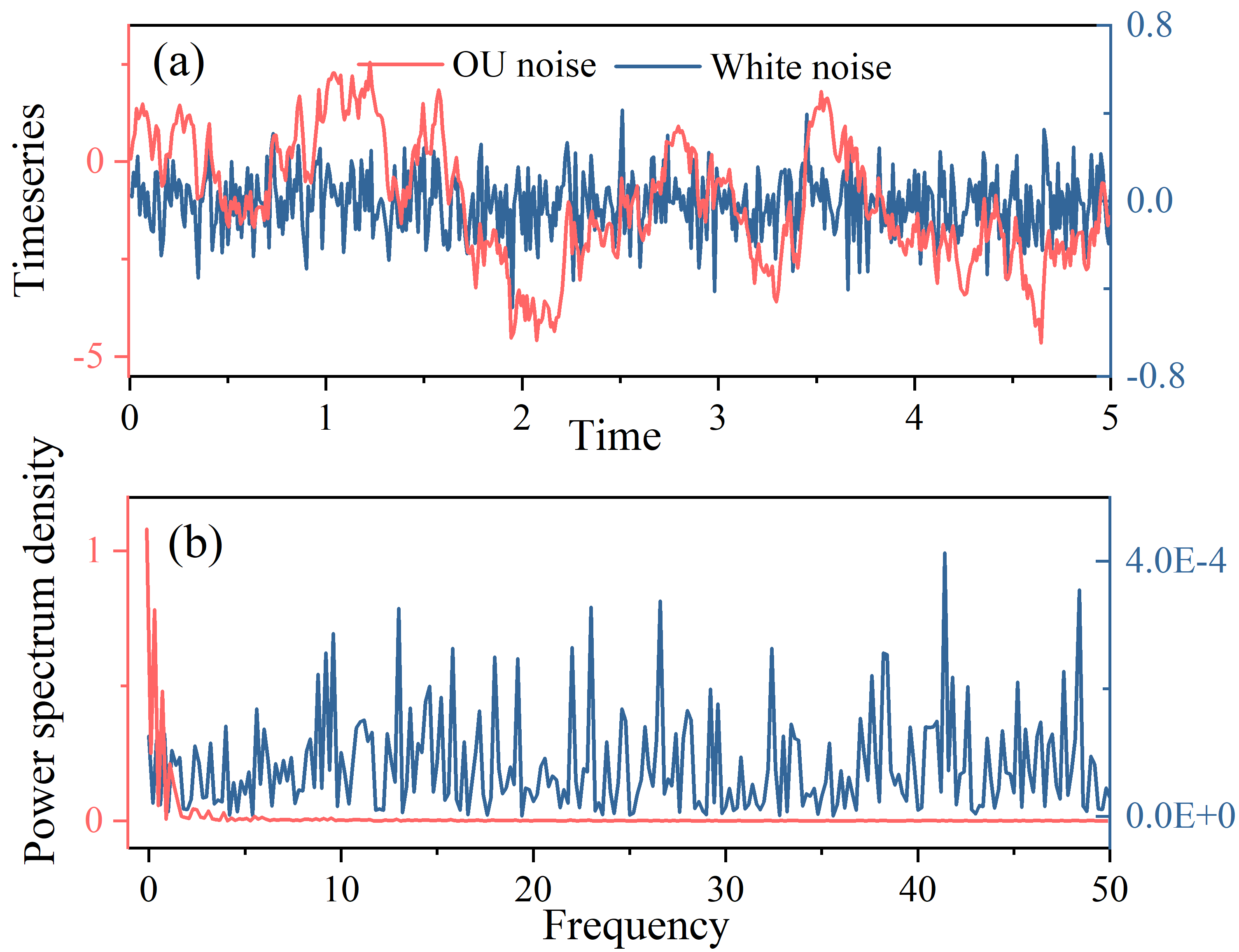}\caption{(a) Timeseries and (b) the corresponding power spectrum density of the Gaussian white noise and the Ornstein-Uhlenbeck noise. For the Gaussian white noise, all the frequencies of its power spectrum exist and pose equal weight. As a result, one need an infinite amount of energy to generate white noise. While for a Ornstein-Uhlenbeck noise, the dependence of power spectrum on frequency manifests a Lorenz function, which suggests that the power of the Ornstein-Uhlenbeck noise concentrates on the low frequency band.}
\label{Fig:1}
\end{figure}

\emph{Mean-field coupled overdamped bistable oscillators.}--The globally coupled (also known as mean-field) overdamped bistable oscillators were primordially proposed by Kometani and Shimizu, in the year of 1975, to investigate the dynamics of muscle contraction~\cite{Kometani1975}. Later on, such a model was extended by Desai and Zwanzig to study the behavior of noise-induced phase transition~\cite{Desai1978}. Nowadays, the globally coupled bistable oscillators have been a popular model in statistical physics and complex systems~\cite{Zagli2021,Liang2021,Liu2023,Liu2024,Tessone2006}, and the dynamics of such model driven by both the weak external signal and the OU noise can be described by the differential equations
\begin{equation}\label{eq:Bistable}
\dot{x_{i}}=x_{i}-x_{i}^{3}+\frac{c}{N}\sum_{j=1}^{N}(x_{j}-x_{i})+A\sin(\omega t)+\zeta_{i}(t),i=1,...N,
\end{equation}
in which $x_{i}$ and $c$ are the state variable of the $i$-th oscillator and the strength of the global coupling, respectively. The sinusoidal 
 term stands for the weak external driving, where $A$ and $\omega$ are the amplitude and frequency. The symbol $\zeta_{i}$ is the OU noise adding on the $i$-th oscillator. The first and second moments of the noise are 
\begin{equation}\label{momentsofOU}
\langle \zeta_{i} \rangle=0,\langle \zeta_{i}(t)\zeta_{j}(t') \rangle=\delta_{ij}\frac{D}{\tau}\exp\left(-\frac{\mid t-t'\mid}{\tau}\right),
\end{equation}
respectively, where $\tau$ and $D$ are the correlation time and noise intensity.
As Fig.~\ref{Fig:1} shows, for a Gaussian white noise $\xi(t)$, the power spectrum, $S_{\textrm{White}}=\int_{-\infty}^{\infty}\langle\xi(t)\xi(t')\rangle e^{-i\omega t}dt=2D$, pervades the whole frequency band. While for a OU noise, the power spectrum $S_{\textrm{OU}}=\int_{-\infty}^{\infty}\langle\zeta(t)\zeta(t')\rangle e^{-i\omega t}dt=2D/(1+\tau^{2}\omega^{2})$, is bounded. Conventionally, the non-Markovian one-dimensional evolution of Eq.~(\ref{eq:Bistable}) can be recast into the Markovian bidimensional process $(x_{i},y_{i})$, which can be further depicted by the Langevin equation~\cite{Hanggi1994,Hanggi1993,Jung1987} 
\begin{eqnarray}\label{eq:Bi-dimensional}
\dot{x_{i}}&=&x_{i}-x_{i}^{3}+\frac{c}{N}\sum_{j=1}^{N}(x_{j}-x_{i})+A\sin(\omega t)+y_{i}\nonumber\\
\dot{y_{i}} &=&-\frac{y_{i}}{\tau} +\frac{\xi_{i}(t)}{\tau},i=1,...N,
\end{eqnarray}
where $\xi_{i}(t)$ are Gaussian random variables with $\langle\xi_{i}(t)\rangle=0$ and $\langle \xi_{i}(t)\xi_{j}(t') \rangle=2D\delta_{ij}\delta(t-t')$.

\begin{figure}
\includegraphics[width=0.9\linewidth]{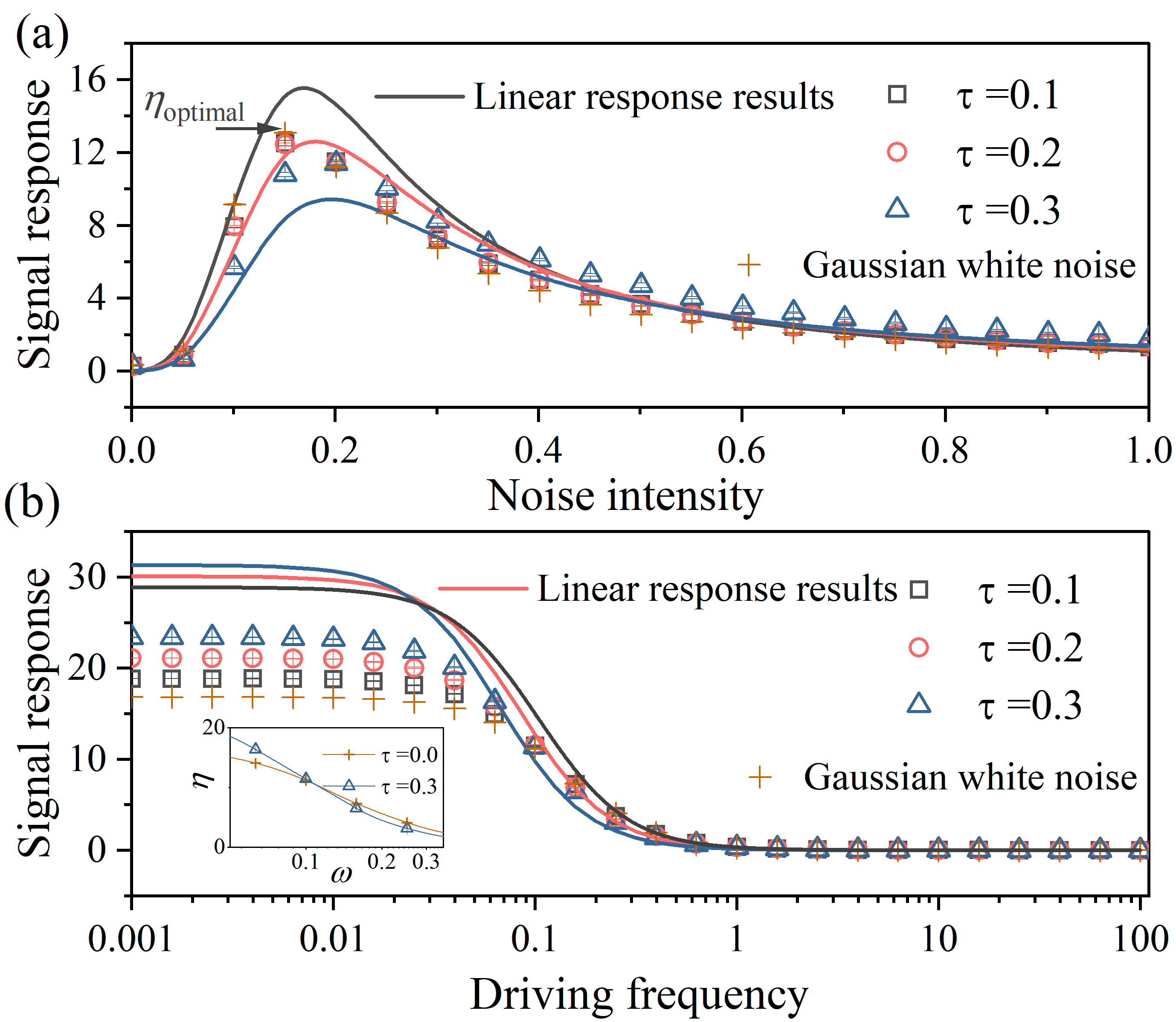}\caption{(a) Collective signal response of mean-field coupled overdamped bistable oscillators versus noise intensity for different correlation times. The amplitude and frequency of the external driving are $A=0.1$ and $\omega=0.1$, respectively. (b) The dependence of signal response on the external driving force for distinct correlation times. The noise intensity is $D=0.2$. The global coupling is fixed to $0.01$. The comparisons of signal response between the Gaussian white noise and OU noise during a narrow frequency region are shown in the inset. }
\label{Fig:2}
\end{figure}

In the present work, we calculate the process of Eq.~(\ref{eq:Bi-dimensional}) by the Euler method with the time step $dt=0.01$, and utilize the spectral amplification factor~\cite{Jung1991,Tessone2006,Liu2023,Liu2024}, 
\begin{equation}\label{eq:Amplification factor}
\eta=\frac{4}{A^{2}} \left| \langle e^{i\omega t}X(t)\rangle \right|^{2}, 
\end{equation}
to measure the collective signal response. The symbol $X(t)=\frac{1}{N}\sum_{j=1}^{N}x_{j}$ denotes the average motion or oscillation of the system and the term $ \langle ...\rangle$ stands for the time average. Additionally, to comprehensively investigate the relationship between the magnitude of resonant signal response and the noise color, we here consider the optimal signal response, $\eta_{\textrm{optimal}}$, as the resonance magnitude~\cite{Liu2023,Liu2024}. In more detail, for each correlation time, we first calculate the signal response for a wide enough range of the noise intensity, say, $D\in[0,1.0]$, then find out the optimal signal response as $\eta_{\textrm{optimal}}$, see Fig.~\ref{Fig:2} (a). When the resonant response is significant, the optimal signal response will be a large value, and vice versa, it is a small constant. Thus, the optimal signal response roughly measures the resonance magnitude~\cite{Hanggi1993,Liu2023,Liu2024}. For clarity, we consider the viewpoint focusing purely on the signal response for a fixed noise intensity as a local view and paying attention to the resonance magnitude as a comprehensive perspective. The results of both the collective signal response and the optimal signal response shown in the present work are averaged over 100 independent realizations.


\begin{figure}
\includegraphics[width=0.9\linewidth]{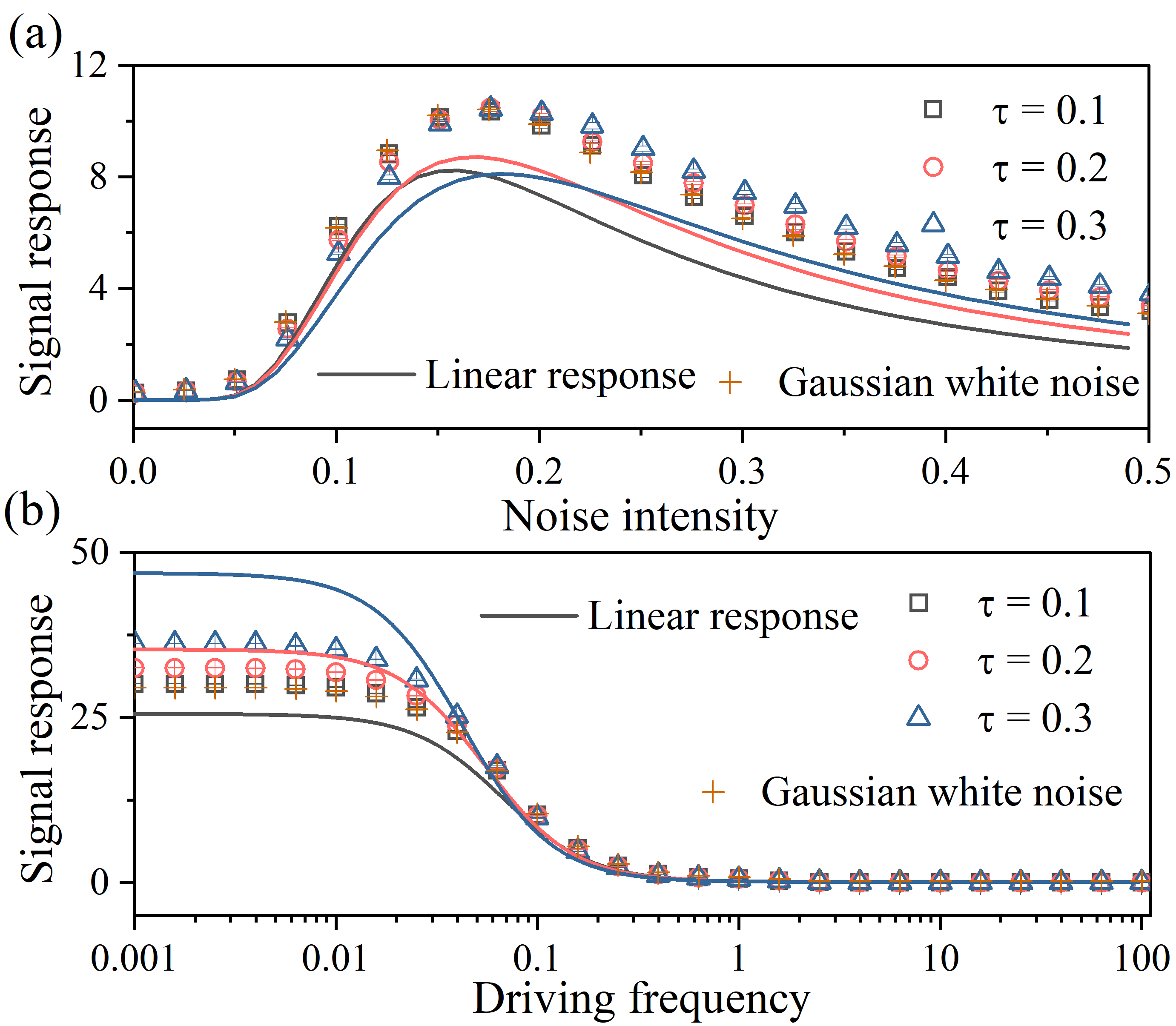}\caption{(a) Signal response of an ensemble of Duffing oscillators versus noise intensity for different correlation times. The amplitude and frequency of the external driving are $A=0.1$ and $\omega=0.1$, respectively. The damping coefficient is $\gamma=0.6$. (b) The dependence of signal response on the external driving force for distinct correlation times. The noise intensity is $D=0.15$. The global coupling is fixed to $0.01$. The results of collective signal response are averaged over $100$ independent realizations.}
\label{Fig:4}
\end{figure}

As Fig.~\ref{Fig:2} (a) shows, with the increase of the noise intensity, the collective signal response showcases a bell-shaped curve for a given correlation time, which manifests the typical characteristic of stochastic resonance. Furthermore, compared to the signal response in the white noise situation (namely $\tau=0$), the optimal signal response and the corresponding optimal noise intensity in the OU noise case are reduced and increased, respectively. These results agree with the conventional conclusion that the additive OU noise weakens the magnitude of SR~\cite{Hanggi1993,Nozaki1999prl,Nozaki1999pre,Mondal2018,Soma2003,Gammaitoni1989,Cabrera1999}. However, such a conclusion can be inverted by adjusting the external driving frequency for a fixed noise intensity. As shown in Fig.~\ref{Fig:2} (b), the collective signal responses for different correlation times behave as inverse-Sigmoid-curves as the increase of the external driving frequency. Interestingly, these curves have an intersection, see the inset in Fig.~\ref{Fig:2} (b), more specifically, when the driving frequency is less than the critical frequency (the frequency corresponds to the intersection), the color of noise improves the signal response, whereas if the frequency is larger than the critical frequency, the noise color weakens it. Nevertheless, a more comprehensive picture displays that the resonance is weakened by the noise color. As Figs.~\ref{Fig:4} (a)-(c) shows, the optimal signal response reduces monotonically as the rise of the noise color. These results suggest that despite for a fixed noise (a local view), the colored noise outperforming the white noise in amplifying a weak low-frequency signal, the whole resonance is impaired by the noise color. 

To put these findings on a solid theoretical ground, we utilize the linear response theory and the effective Fokker-Planck approximation~\cite{Hanggi1993,Liu2019} to deduce the collective signal response,
\begin{eqnarray}\label{eq:Linearresponse}
\eta_{\textrm{over}}=\left|\chi(\omega)\right|^{2}=\frac{\left[a(1-ca)\lambda_{\textrm{min}}^{2}\right]^{2}+(a\omega\lambda_{\textrm{min}})^{2}}{\left[(1-ca)^{2}\lambda_{\textrm{min}}^{2}+\omega^{2}\right]^{2}},
\end{eqnarray}
where $a=(\langle x^{2}\rangle_{0}+\tau D)/D$, the symbol $\lambda_{\textrm{min}}=[\sqrt{2}(1-3\tau/2)/\pi]\exp(-1/4D)$, the term $\langle x^{2} \rangle_{0}$ denotes the stationary average of the unperturbed process, see ~\cite{SMliucong2024} for the details of the theoretical signal response.

As Fig.~\ref{Fig:2} (a) shows, taking $\tau=0.1$ as an example, the theoretical signal response depicts as a bell-shaped curve with the rise of the noise intensity, which coincides with the corresponding numerical results. Besides, as the noise color increases, an explicit reduction on the signal response can be viewed. On the other hand, for a fixed noise intensity, the theoretical signal response of Eq.~(\ref{eq:Linearresponse}) monotonously decreases as the increase of the driving frequency manifesting the typical inverse-Sigmoid-curve. For different correlation times, these theoretical inverse-Sigmoid-curves have an decussation at an intermediate frequency as well, see Fig.~\ref{Fig:2} (b), which demonstrates that the colored noise can perform a better role than white noise in amplifying signal response in a specific circumstance. 

Based on the definition of $\eta_{\textrm{optimal}}$ and the signal response of Eq.~(\ref{eq:Linearresponse}), we can expediently obtain the optimal signal response through the numerical method concerning the maximum value. As Figs.~\ref{Fig:4} (a)-(c) shows, the optimal signal response obtained by the linear response theory depicts the monotonously decreasing trend as the increase of the noise color. Such a result is robust to the driving frequency.
These results further verify that the color of the additive OU noise weakens the magnitude of SR in coupled overdamped situation. 

Moreover, if we consider that the optimal signal response emerges when the escape rate is twice the external driving frequency, i.e., $\lambda_{\textrm{min}}/\pi=2\omega$, we can obtain the optimal noise intensity corresponding to the optimal signal response as 
\begin{equation}\label{eq:optimalD}
D_{\textrm{optimal}}=\frac{1}{4}\left[ \ln\left( \frac{1-\frac{3}{2}\tau}{\sqrt{2}\omega}\right)\right]>D_{\textrm{optimal}}(\tau=0),    
\end{equation}
and the optimal signal response is directly proportional to $1/(2c\tau+3)D_{\textrm{optimal}}^{2}$. These results demonstrate that the noise color suppresses the escaping behavior of the bistable oscillators and thereby weakens the resonance, see ~\cite{SMliucong2024} for the details.
\begin{figure}
\includegraphics[width=1.0\linewidth]{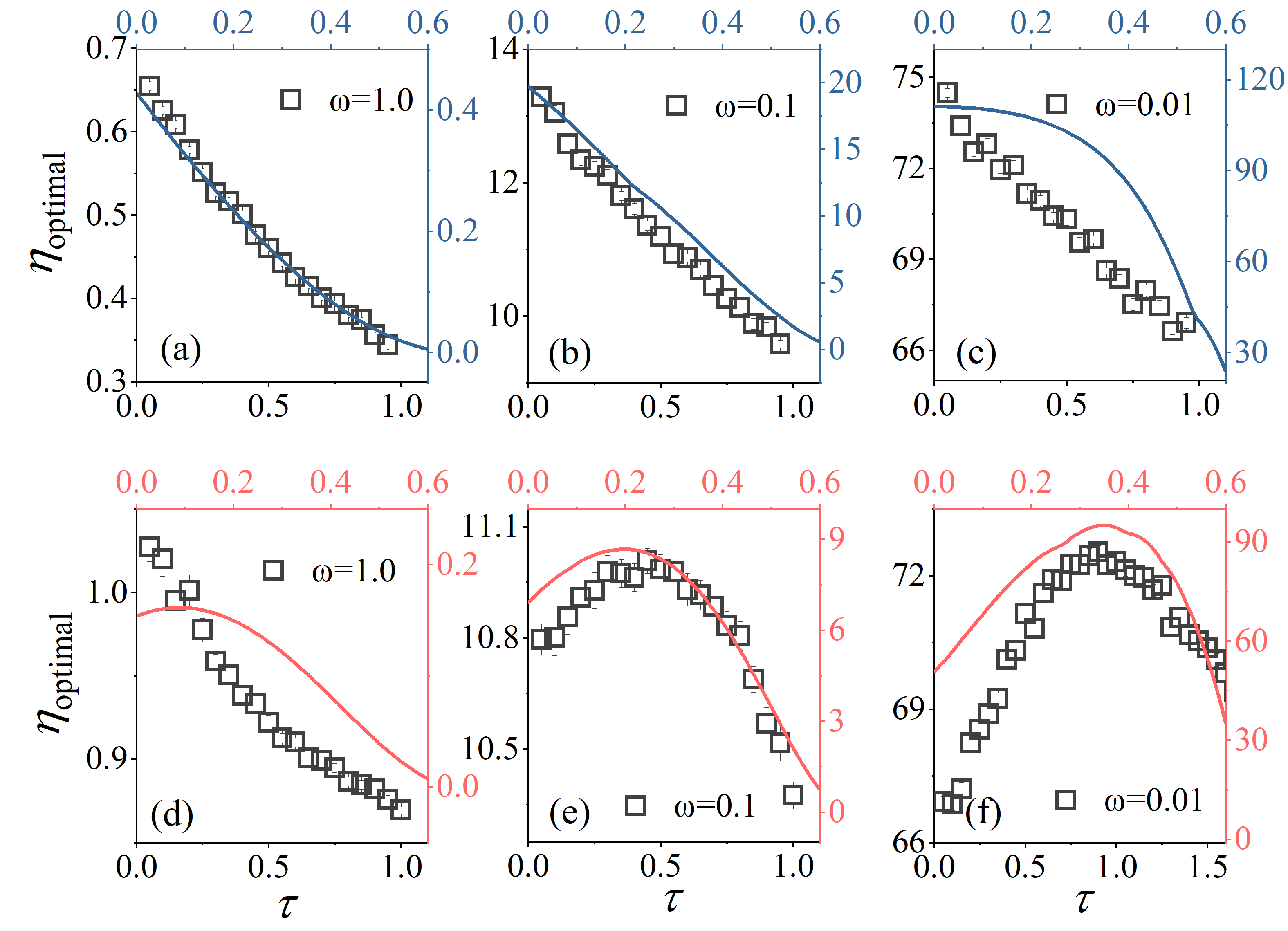}\caption{ The relationship between the resonance magnitude and the noise color (correlation time). (a)-(c) The optimal signal response of mean-field coupled overdamped bistable oscillators versus correlation time for three driving frequencies. (d)-(f) The optimal signal response of an ensemble of Duffing oscillators versus correlation time for three different driving frequencies. The solid lines denote the optimal signal responses obtained by the linear response theory. The amplitude and frequency of the external driving are $A=0.1$ and $\omega=0.1$, respectively. The damping coefficient is $\gamma=0.6$.}
\label{Fig:3}
\end{figure}



\emph{Mean-field coupled underdamped Duffing oscillators.}--Since the Duffing oscillator can not only depict the bistable feature in fields spanning from biophysics to engineering but also character the damping effect~\cite{Peters2021,Cuairan2022,Meucci2016,Liu2019,Liang2021}, here we consider the mean-filed coupled Duffing oscillators driven by both the weak force and the OU noise to explore the influence of the noise color on SR, of which the dynamics can be represented by the differential equations, 
\begin{eqnarray}\label{eq:Duffingoscillators}
\ddot{x}_{i}+\gamma\dot{x}_{i}&=&x_{i}-x_{i}^{3}+\frac{c}{N}\sum_{j=1}^{N}(x_{j}-x_{i})\nonumber\\
&+&A\sin(\omega t)+\zeta_{i}(t),i=1,...,N,   
\end{eqnarray}
in which the constant $\gamma$ is the damping coefficient. The symbol $\zeta_{i}(t)$ is still the OU noise with zero mean and the self correlation, 
\begin{equation}
\left\langle\zeta_{i}(t)\zeta_{j}(t')\right\rangle=\delta_{ij}\frac{D\gamma}{\tau}\exp\left(-\frac{\mid t-t'\mid}{\tau}\right).   
\end{equation}

Similarly, the non-Markovian one dimensional evolution of Eq.~(\ref{eq:Duffingoscillators}) can be recast into the Markovian three dimensional process $(v_{i},x_{i},y_{i})$ as 
\begin{eqnarray}\label{eq:Tridimensionalduffing}
v_{i}&=&x_{i}-x_{i}^{3}-\gamma v_{i}+\frac{c}{N}\sum_{j=1}^{N}(x_{j}-x_{i})+A\sin(\omega t)+y_{i}\nonumber\\
\dot{x_{i}}&=&v_{i}; \dot{y_{i}} =-\frac{y_{i}}{\tau} +\frac{\xi_{i}(t)}{\tau},i=1,...N,
\end{eqnarray}
where $\xi_{i}(t)$ are also Gaussian random variables with $\langle\xi_{i}(t)\rangle=0$ and $\langle \xi_{i}(t)\xi_{j}(t') \rangle=2D\gamma\delta_{ij}\delta(t-t')$.

As Fig.~\ref{Fig:3} (a) shows, for a fixed correlation time, a clear bell-shaped signal response curve can be viewed as the increase of the noise intensity. Furthermore, compared to the signal response in the white noise situation, taking $\tau=0.3$ as an example, the signal response presents a slight enhancement, especially in the range of $D\in[0.2,0.5]$. On the other side, for a given noise intensity, the signal response decreases monotonously as the driving frequency gradually rises, and if the driving frequency is significantly low, i.e., $\omega=0.01$, the explicit signal response enhancement induced by the noise color can be observed, see in Fig.~\ref{Fig:3} (b). These results indicate that the noise color plays a positive role in enhancing signal response in a local viewpoint. Moreover, from a comprehensive viewpoint, the resonance magnitude decreases as the noise color rises for a high frequency driving signal, as shown in Fig.~\ref{Fig:4} (d) . While for an intermediate or low frequency driving force, the optimal signal response rises first and goes down gradually as the noise color increases, which manifests a resonance-like SR enhancement, see Figs.~\ref{Fig:4} (e) and ~\ref{Fig:4} (f). These results suggest that the color of the OU noise can not only enhances the SR but also optimizes it at an intermediate value.

Resemble the approach in the overdamped situation, we can obtain the theoretical signal response of the Duffing oscillators on the basis of the linear response theory as 
\begin{equation}\label{eq:finalresponse}
\eta_{\textrm{under}}=\frac{\left[\lambda_{\textrm{min}}\gamma\langle x(t)^{2}\rangle_{0}b\right]^{2}}{[1-c\gamma \langle x(t)^{2}\rangle_{0} b]^{2}\lambda_{\textrm{min}}^{2}+\omega^{2}},
\end{equation}
in which $b=[1-\tau(1-3\langle x^{2}\rangle_{0})]/D$.  See also~\cite{SMliucong2024} for the details of the theoretical signal response.

As Fig.~\ref{Fig:3} (a) shows, for a given $\tau$, the theoretical signal response captures an explicit bell-shaped enhancement as the noise intensity rises, which reasonably agrees with the numerical result. Moreover, the signal response of Eq.~(\ref{eq:finalresponse}) decreases in an inverse-Sigmoid-curve as the external driving frequency increases for a fixed correlation time, see in Fig.~\ref{Fig:3} (b). In addition, we can also obtain the theoretical optimal signal response on the basis of the Eq.~(\ref{eq:finalresponse}), as shown in Figs.~\ref{Fig:4} (a)-(c), for a high-frequency external signal, the theoretical resonance magnitude undergoes an platform and then decreases as the noise color increases. Whereas for an intermediate or low frequency driving force, the theoretical magnitude of SR rises firstly and then decreases as the correlation time increases, which aligns with the numerical result that the noise color can nonmonotonically enhance SR.
These results confirm that the color of the additive OU noise can play a positive role in enhancing SR.

Similarly, we can obtain the optimal signal response approximates $1/f(D_{\textrm{optimal}})$ based on Eq.~(\ref{eq:optimalD}), in which $f(D_{\textrm{optimal}})=[(1+0.25\pi^{2})/\gamma^{2}(1+2\tau)^{2}]D_{\textrm{optimal}}^{2}-[2c/\gamma(1+2\tau)]D_{\textrm{optimal}}+c^{2}$. The axis of the symmetry for $f(D_{\textrm{optimal}})$ moves right toward as the noise color increases, which evokes the nonmonotonic growing interval. As a result, the complicated interactions among damping, coupling, and the noise color are responsible for the optimization behavior, see ~\cite{SMliucong2024} for the details. 

Despite these theoretical results are qualitatively in line with the numerical ones (also the overdamped situation), the significant deviations can be viewed from a quantitative perspective in Fig.~\ref{Fig:4}. Two reasons are mainly responsible for these deviations. On the one hand, the effective Fokker-Planck approximation holds good only for small correlation time. On the other hand, we roughly consider the Kramers escape rate instead of the averaged self-correlation, which causes the signal response deviation.

\emph{Functional significance of colored noise.}--Since colored noise is encountered in a wide variety of natural and manmade systems and has been recognized as an emblematic feature of complexity~\cite{Gilden1995}, The question concerning the functional significance of colored noise attracts research interests in different directions. Particularly, Nozaki and coworkers investigated theoretically and experimentally the role of the noise color on SR, they found that the $1/f$ noise (one typical type of colored noise, of which the power spectrum showcases the inverse-proportional function character) can be better than the white noise for enhancing the signal response of a rat skin neuron from a local version, say, the signal response of the $1/f$ noise situation can be larger than that of the white noise one for a given noise intensity~\cite{Nozaki1999prl,Nozaki1999pre}. Nevertheless, for a comprehensive version, the optimal signal response is reduced as the noise color increases. Such results confirm the conventional sense that the additive colored noise has no functional advantage in enhancing the collective signal response~\cite{Hanggi1993,Nozaki1999prl,Nozaki1999pre,Mondal2018,Soma2003,Gammaitoni1989,Cabrera1999}. Our results concerning the mean-field coupled overdamped bistable oscillators coincide with these conclusions. Nevertheless, the main results in the underdamped case in the present work reveal that colored noise can perform a better role than white noise in enhancing the signal response in both local and global viewpoints. These results go against the conventional sense that the additive colored noise impairs the magnitude of stochastic resonance, thus giving a clue that colored noise may has functional significance in weak signal amplification in coupled bistable elements. 

\emph{Conclusion.}--In conclusion, we have theoretically and numerically investigated the collective signal response of the mean-field coupled overdamped and underdamped bistable oscillators with respect to the additive OU noise and the weak periodical signal, and given the signal response formula embracing the influence of both the noise intensity and the driving frequency for the two bistable systems based on the linear response theory. Moreover, we demonstrated that (I) for the overdamped bistable oscillators, from a local viewpoint, like a given noise intensity, the ensemble signal response in the OU noise case can be better than that of the white noise one. While for a comprehensive version, the SR magnitude is gradually reduced as the noise color increases. (II) More interestingly, for the underdamped bistable oscillators, from both the local and comprehensive viewpoints, the noise color can strengthen the stochastic resonance. Specifically, when the frequency of the external driving signal is not large, the resonance magnitude can be nonmonotonically improved as the noise color increases.

Despite the large interest in characterizing the functional significance of colored noise, such a theme has not been fully elucidated. Our results benefit the understanding of the functional significance of colored noise from the perspective of the weak signal response. Besides, as both the Duffing equation and the colored noise are common in mechanics and engineering, our results, with the promising applications in engineering like the micro-electromechanical energy harvester, await experimental confirmation~\cite{Peters2021,Mondal2018}.

\emph{Acknowledgments.}--Cong Liu warmly thanks professors Xiaoming Liang and Xiyun Zhang for the insightful suggestions.  
This work is funded by Science Research Project of Hebei Education Department under Grant No. QN2025065 and Science Foundation of Hebei Normal Universty under Grant No. L2025B10.
We acknowledge financial support from the National Natural Science Foundation of China (Grants No. 12375032 and No. 12247101), and from the Fundamental Research Funds for the Central Universities (Grant No. lzujbky-2023-ey02).


\end{document}